\documentclass[conference]{IEEEtran}
\IEEEoverridecommandlockouts
\usepackage{cite}
\usepackage{amsmath,amssymb,amsfonts}
\usepackage{algorithmic}
\usepackage{graphicx}
\usepackage{textcomp}
\usepackage{xcolor}
\usepackage{hyperref}
\usepackage{caption}
\usepackage{subcaption}
\usepackage{multirow}
\usepackage{placeins}
\usepackage{comment}
\def\BibTeX{{\rm B\kern-.05em{\sc i\kern-.025em b}\kern-.08em
    T\kern-.1667em\lower.7ex\hbox{E}\kern-.125emX}}
\DeclareUnicodeCharacter{2212}{-}

\FloatBarrier
\begin{document}

\title{Transformer-based de novo peptide sequencing for data-independent acquisition mass spectrometry\\
%{\footnotesize \textsuperscript{*}Note: Sub-titles are not captured in Xplore and should not be used}
\thanks{This work was supported by the National
Library of Medicine of the National Institutes of
Health under award number R15LM013460, and
the National Center for Complementary \&
Integrative Health of the National Institutes of
Health under the award number R01AT011618
https://www.nih.gov/.}
}

\author{\IEEEauthorblockN{Shiva Ebrahimi}
\IEEEauthorblockA{\textit{Computer Science \& Engineering} \\
\textit{University of North Texas}\\
Denton, USA \\
Shiva.Ebrahimi@unt.edu}
\and
\IEEEauthorblockN{Xuan Guo}
\IEEEauthorblockA{\textit{Computer Science \& Engineering} \\
\textit{University of North Texas}\\
Denton, USA \\
Xuan.Guo@unt.edu}
}

\maketitle

\begin{abstract}
Tandem mass spectrometry (MS/MS) stands as the predominant high-throughput technique for comprehensively analyzing protein content within biological samples. This methodology is a cornerstone driving the advancement of proteomics. In recent years, substantial strides have been made in Data-Independent Acquisition (DIA) strategies, facilitating impartial and non-targeted fragmentation of precursor ions. The DIA-generated MS/MS spectra present a formidable obstacle due to their inherent high multiplexing nature. Each spectrum encapsulates fragmented product ions originating from multiple precursor peptides. This intricacy poses a particularly acute challenge in de novo peptide/protein sequencing, where current methods are ill-equipped to address the multiplexing conundrum. In this paper, we introduce Transformer-DIA, a deep-learning model based on transformer architecture. It deciphers peptide sequences from DIA mass spectrometry data. Our results show significant improvements over existing STOA methods, including DeepNovo-DIA and PepNet. Transformer-DIA enhances precision by 15.14\% to 34.8\%, recall by 11.62\% to 31.94\% at the amino acid level, and boosts precision by 59\% to 81.36\% at the peptide level. Integrating DIA data and our Transformer-DIA model holds considerable promise to uncover novel peptides and more comprehensive profiling of biological samples. Transformer-DIA is freely available under the GNU GPL license at \url{https://github.com/Biocomputing-Research-Group/Transformer-DIA}.

\end{abstract}

\begin{IEEEkeywords}
De novo peptide sequencing, transformer, data-independent acquisition (DIA), mass spectrometry
\end{IEEEkeywords}

\section{Introduction}

In proteomics, liquid chromatography combined with tandem mass spectrometry (LC-MS/MS) is a widely used method for identifying and quantifying proteins~\cite{mccormack1997direct}. In the traditional data-dependent acquisition (DDA) approach, proteins are first broken down into smaller units called peptides, which are then measured using a mass spectrometer. The mass spectrometer selects certain peptides based on their signal strength in a mass spectrum (MS1) and breaks them down further to create tandem mass spectra (MS2). Researchers then try to match these MS2 spectra with entries in a protein database . However, this method has limitations when it comes to consistently identifying and quantifying peptides that are present in low amounts. An alternative technique called data-independent acquisition (DIA) involves intentionally fragmenting all peptides within a set of sliding mass-to-charge windows (e.g., 12 m/z). This approach results in highly accurate and reproducible data for peptide identification and quantification~\cite{fernandez2020impact}. Nevertheless, DIA data presents its own set of challenges. However, the broad mass-to-charge range may reduce the method's ability to select specific peptides \cite{hunter2023perspectives}, and dealing with complex data from multiple peptides can be intricate. Thus, accurately identifying peptides and proteins in DIA data using computational methods poses significant challenges for researchers.

The process of translating MS/MS spectra into peptides primarily encompasses two approaches: (1) the utilization of a database search engine, which relies on databases containing known sequences, and (2) de novo peptide sequencing, a method that deciphers MS/MS spectra without database searching or prior amino acid sequence knowledge. Recently, various de novo peptide sequencing algorithms have been developed. These algorithms mainly employ one of the following approaches: i) graph-based methods, and ii) neural networks. The first proposed de novo peptide sequencing algorithm computes mass differences between observed spectra and all possible predicted peptides and define scoring functions to identify the best-matched peptide sequence~\cite{beslic2023comprehensive}. Graph-based de novo peptide sequencing is a method that interprets mass spectrometry data by constructing a directed acyclic graph (DAG) to identify potential amino acid sequences~\cite{bartels1990fast, yan2011applications}. The graph nodes of mass peaks are connected with at least one or more amino acid mass difference. In DAG, the optimal peptide sequence was determined by finding the longest path. Dynamic programming is usually applied for reducing the computing time for spectrum graphs. Commercial software such as PEAKS~\cite{ma2003peaks} and Novor \cite{ma2015novor} are dynamic programming-based algorithms that applied probabilistic scoring functions to select the best peptide sequences. One obvious drawback of graph-based and dynamic programming-based algorithms is the time complexity that increases dramatically with the high spectrum resolution generated by advanced spectrometers.

Recently, many de novo peptide sequencing algorithms based on neural networks have shown success in peptide identification. For example, Tran et al. proposed DeepNovo that is the first neural network-based method~\cite{tran2017novo}. It employs a combination of convolutional neural networks (CNN) and long short-term memory (LSTM) networks within its architecture to interpret mass spectrometry data for de novo peptide sequencing. Qiao et al. designed PointNovo that incorporates an invariant network structure (T-Net) with RNN while using knapsack dynamic programming to optimize search space~\cite{qiao2021computationally}. DeepNovo-DIA was developed as the extension of DeepNovo model \cite{tran2017novo} for DIA MS/MS spectra by employing LSTM to learn peptide language and CNN to understand high-dimensional fragment ions and their correlation with precursor ions~\cite{tran2019deep}. Liu et al. developed PepNet, which is based on convolutional neural networks for both DIA and DDA data~\cite{liu2022pepnet}. Li et al. proposed DpNovo that combines two deep learning models for feature learning and dynamic programming for determining optimal amino acid sequences, ultimately providing predicted sequences with mass values representing uncertain mass intervals~\cite{li2023dpnovo}. Vertens et al. devised SMSNet that introduced a hybrid de novo peptide sequencing and database-search approach by which the predicted amino acid sequences were searched in the database and replaced with the best match~\cite{karunratanakul2019uncovering}. Yang et al. developed pNovo3, which combines a learning-to-rank framework with a deep learning model to distinguish similar peptide candidates for each spectrum~\cite{yang2019pnovo,zhou2017pdeep}. Yilmaz et al. designed Casanovo that uses a transformer framework to map directly from a sequence of observed peaks to a sequence of amino acids~\cite{yilmaz2022novo}. Nonetheless, most of the current de novo peptide sequencing models were initially tailored for DDA data, making them unsuitable for direct application to DIA data where an MS2 spectrum comprises product ions from several precursor peptides. Moreover, even the state-of-the-art approaches such as DeepNovo, DeepNovo-DIA, and SMSNet exhibit reduced sensitivity at the peptide level compared to the amino acid level and suffer from inefficient memory usage as they represent each spectrum with an extensive vector for MS intensities. 

In this paper, we present an extension to the Casanovo model, called Transformer-DIA, aimed at translating DIA spectra into peptide sequences. We enhanced the transformer-based model with a novel encoder block designed to integrate information from MS1, MS2, and precursor profile before feeding it into the transformer encoder. We employed three distinct methods for integrating the encoded inputs: concatenation, standard attention mechanisms, and multi-head-attention layers. We assessed the performance of Transformer-DIA using three Homo sapiens DIA datasets released from the DeepNovo-DIA paper~\cite{tran2019deep}. Our experimental results demonstrate that Transformer-DIA consistently outperformed STOA method, including both DeepNovo-DIA and PepNet, in terms of amino acid- and peptide-level precision and recalls.

\section{Methods}

\begin{figure}[htpb]
\centering
\includegraphics[width=\linewidth]{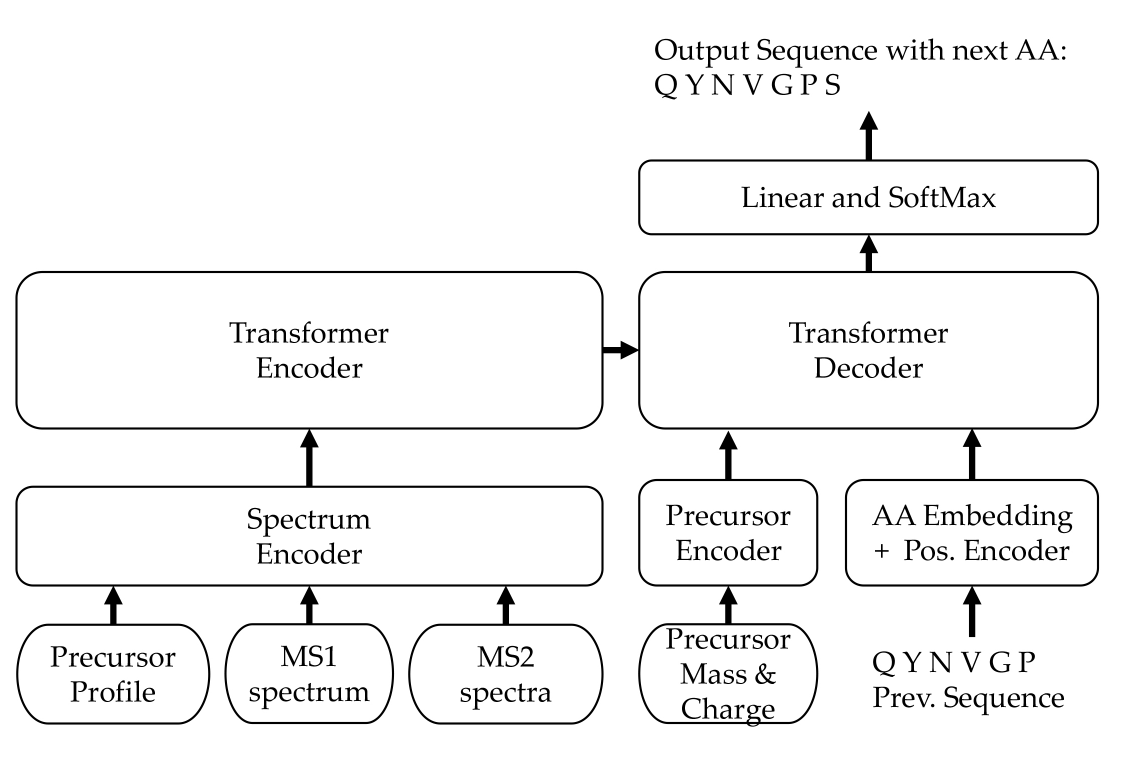}
\caption{Transformer-DIA model architecture with inputs and outputs.}
\label{casanovo-dia}
\end{figure}

To enhance de novo peptide sequencing performance for DIA spectra, we present Transformer-DIA, a transformer-based model outlined in Fig.~\ref{casanovo-dia}. Our key contribution is the expansion of the spectrum encoder block, enabling the learning of essential DIA spectra features. Transformer-DIA comprises a stack of transformer encoders and decoders both equipped with an attention mechanism. Transformer-DIA uses the standard transformer encoders and decoders~\cite{vaswani2017attention} that are responsible for acquiring latent representations of input mass spectra and deciphering the amino acid sequence linked to the originating peptide, respectively. In instead of utilizing a recurrent neural network (RNN), Transformer-DIA employs a positional encoder to grasp the relative position and contextual information within the input data, using self-attention mechanisms to mathematically assess the relationships among sequential data based on assigned relative positions. In de novo peptide sequencing, the relationship between the m/z values of observed peaks in mass spectra holds significance, as these differences signify the presence of amino acids. The decoder leverages encoded precursor peaks, containing vital m/z difference, to predict amino acid sequences. We explored three distinct approaches for combining the encoded inputs: concatenation, standard attention and multi-head-attention mechanisms. Transformer-DIA is freely available under the GNU GPL license at \url{https://github.com/Biocomputing-Research-Group/Transformer-DIA}. Further elaboration on our modeling strategy is provided in the subsequent sections.

%The encoder takes d-dimensional spectrum peak embeddings as input and outputs $d$-dimensional latent representation vectors for each peak. Subsequently, the decoder takes as input these representations of prefix amino acids, coupled with a $d$-dimensional precursor embedding encapsulating precursor m/z and charge information, to predict the next amino acid in the peptide sequence. 

%“Meaning is a result of the relationship between things, and self-attention is a general way of learning relationships” \cite{vaswani2017attention}. 

%The encoder takes the variable length peaks, subsequentially the decoder infers the variable length amino acid sequences. Subsequently, the decoder utilizes encoded peaks containing important information about the m/z difference between peaks to predict amino acid sequences. Casanovo is designed for DDA spectra, so it only considers bag of peaks containing m/z and intensity. 

\begin{figure*}[htb!]
\centering
\includegraphics[width=0.95\linewidth]{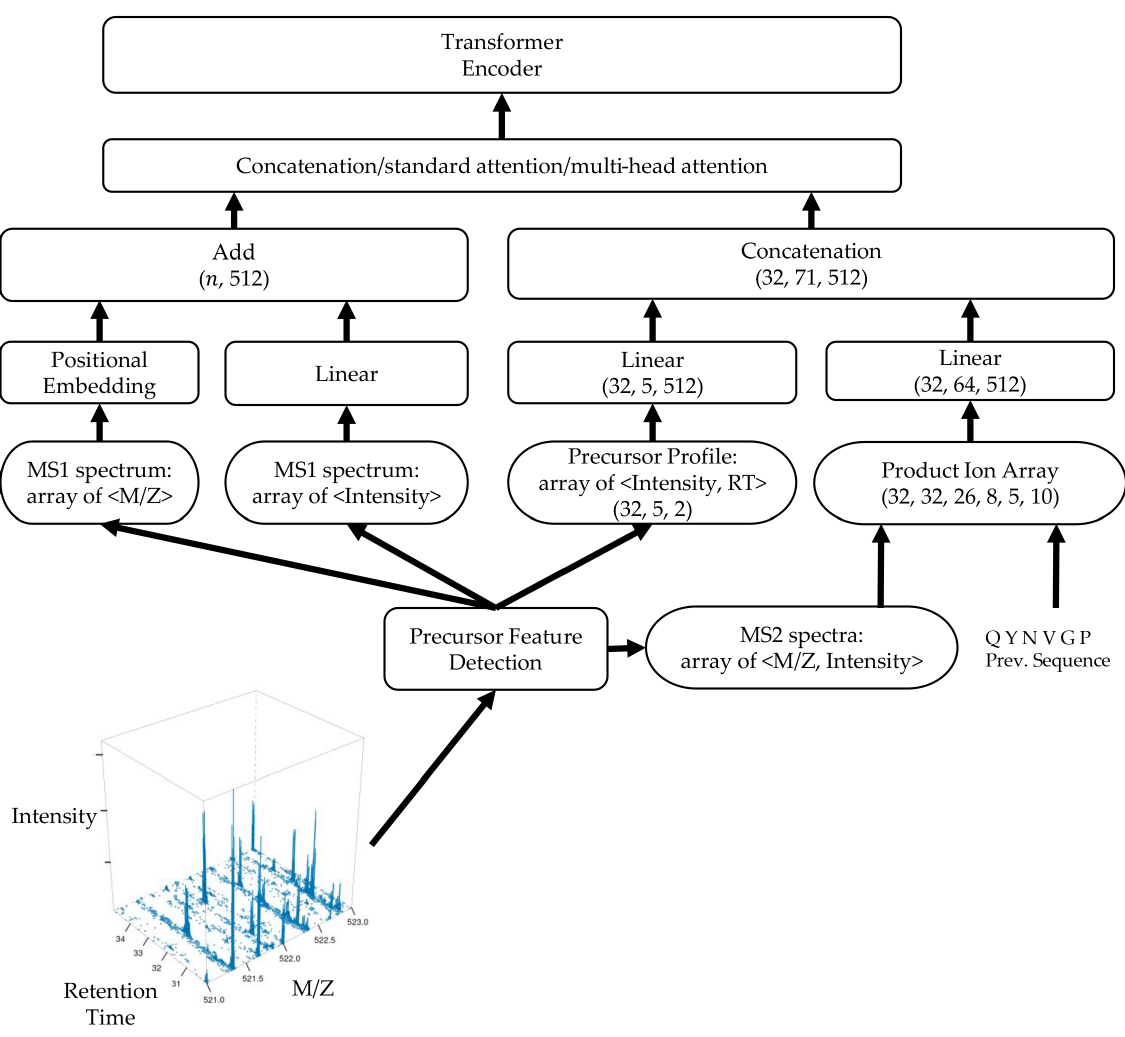} 
\caption{The architecture of Transformer-DIA spectrum encoder.}
\label{encoder-block}
\end{figure*}

%: To feed the encoded spectra to the transformer encoder, we integrate the results of encoded peaks, MS1, and MS2. The set of peaks (m/z, I) of each spectrum is encoded using sinusoidal functions for m/z values and a linear layer for intensity values \cite{yilmaz2022novo}. The output of peak encoder is the summed m/z and intensity embedding used as the input of cat function. For encoding each of MS1 and MS2, we simply applies a linear layer. Finally, three encoded vector are concatenated to lunch to the transformer encoder.

\subsection{Spectrum Encoder in Transformer-DIA}

the peptide level while ensuring scalability 
This enhancement in performance, coupled with its scalability and comparable runtime efficiency, positions Casanova-DIA as a promising tool for advancing peptide sequencing in the field of LC MS/MS DIA data analysis.  into three training, validation, and testing sets with a ratio of 0.95, 0.5, and 0.5. Thus, the DIA data sets were annotated using the PEAKS DB database. Each spectrum is assigned a grand truth peptide sequence as the label. We also applied cross-validation without having any shared peptide sequence. The experimental results show that Transformer-DIA outperforms both DeepNovo-DIA and PepNet at both the amino acid level and significantly at the peptide level.

The Spectrum Encoder provides contextualized information for the Transformer Encoder, as depicted in Figure~\ref{encoder-block}. It accepts three primary inputs: an MS1 spectrum, a precursor profile, and a product ion array. In the context of LC-MS/MS DIA data, we adopted a consistent preprocessing approach similar to that utilized in the DeepNovo-DIA method to gather these essential inputs. To elaborate, we employed the peak caller from the reference~\cite{zhang2012peaks} to generate a comprehensive list of precursor features. This list encompassed crucial details such as m/z values, charge states, retention-time centers, and intensity values spanning the retention-time range. The latter two parameters were instrumental in defining the characteristics of the precursor profile. In Figure~\ref{encoder-block}, we represent a tabulation of $(Intensity, RT)$ pairs constituting the precursor profile as a tensor with dimensions $(32, 5, 2)$, where the numeric value 32 corresponds to the batch size and the numeric value 5 corresponds to five precursor peaks close to  m/z and retention-time centers. For each identified precursor feature, we systematically compiled a set of MS2 spectra that fell within the specified retention-time range while ensuring their DIA m/z windows adequately covered the feature's m/z. Additionally, we included the MS1 spectrum closest in proximity to the retention-time center of the respective feature. Regarding the retention-time dimension of the precursor profile, we standardized this to include five spectra, specifically those that exhibited the closest alignment with the feature's retention-time center.  

Each MS2 spectrum was encoded as a vector of size 150,000, where the mass of each product ion corresponded to a unique index, and the intensity of the product ion determined the value at that specific index. The set of five carefully selected MS2 spectra corresponding to a specific feature were collectively organized into a normalized matrix with dimensions $(5, 150,000)$. For the processing of each input prefix, Transformer-DIA generated a product ion array with dimensions $(32, 32, 26, 8, 5, 10)$. These dimensions respectively represented the batch size, the length of a peptide sequence, the number of potential amino acid candidates, the range of product ion types, the count of associated spectra, and the size of the model's analysis window. Notably, we defined eight product ion types, denoted as 'b,' 'y,' 'b(2+),' 'y(2+),' 'b-H2O,' 'y-H2O,' 'b-NH3,' and 'y-NH3.' In tandem with an amino acid candidate, the preceding sequence, and a specific product ion type, Transformer-DIA calculated the mass of the product ion. Subsequently, it identified the product ion's corresponding location within the product ion array. Transformer-DIA proceeded to extract an intensity window with a size of 10 centered around the product ion's identified location. Then we employed two separate linear layers to project both the precursor profile and the product ion array into lower-dimensional vectors, yielding dimensions of $(32, 5, 512)$ and $(32, 64, 512)$, respectively.

%Each MS2 spectrum was represented by a vector of size 150,000, in which the mass of an product ion corresponded to an index and the product ion intensity was the vector value at that index. The five selected MS2 spectra of a feature were stored in a normalized matrix of size $(5, 150000)$. For each input prefix, Transformer-DIA computes a six-dimensional array of shapes $(32, 32, 26, 8, 5, 10)$, where the first dimension is the batch size, the second dimension is the length of a peptide sequence, the third dimension is the number of possible candidates of amino acids, the fourth dimension is the number of product ion types, the fifth dimension is the number of associated spectra, and the sixth dimension is the model’s window size. We use eight product ion types: b, y, b(2+), y(2+ ), b-H2O, y-H2O, b-NH3, and y-NH3. With the amino acid candidate, the previous sequence, and a specific product ion type, Transformer-DIA calculates the product ion mass. Following that, Transformer-DIA identifies the ion's location on the intensity vector, i.e., normalized matrix $(5, 150000)$, of feature-associated spectra. Transformer-DIA then extracts an intensity window of size 10 around the product ion location. We employed two linear layers to project the precusor profile and product ion array to a low dimensional vector with the size of $(32, 5, 512)$ and $(32, 64, 512)$, respectively. 

To embed the MS1 spectrum, we followed a similar methodology as outlined in the Casanovo paper~\cite{yilmaz2022novo}. In essence, a consistent sinusoidal m/z positional function was employed to map each m/z value to a 512-dimensional vector. The intensity information was integrated using a linear layer. To construct an embedding vector with dimensions of $(32, n, 512)$, where $n$ signifies the count of peaks in the MS1 spectrum, we performed an element-wise addition of the m/z and intensity embedding vectors. 

To input multi-encoded DIA spectra into the transformer encoder, we amalgamated the encoded vectors using three distinct approaches: concatenation, standard attention~\cite{luong2015effective}, and multi-head attention~\cite{vaswani2017attention}. In the case of standard attention, we calculated a contextual vector to encapsulate the information from the encoded precursor profile, MS1 spectrum, and MS2 spectra. The attention layer accepts the encoded vectors as inputs, denoted as ${h_{(t,i)}}$, and computes $c_t$ as a new integrated vector, as detailed in Equations~\ref{equ1}, \ref{equ2}, and \ref{equ3}. The context vector $c_t$ generated at time step $t$ represents a weighted summation of $h_t$, where the attentional weights $att_t$ are determined using a softmax layer on the attentional vector $e_t$. This vector $e_t$ is created through the trainable encoded vector $W_c$.

\begin{equation}\label{equ1}
	 c_t = \sum_i att_{t,i} h_{t,i}
\end{equation}

\begin{equation}\label{equ2}
	  att_t = softmax(e_t)
\end{equation}

\begin{equation}\label{equ3}
	  e_t  = tanh{(W_ch_t)}
\end{equation}

In the case of multi-head attention, Transformer-DIA concurrently focuses on detailed information from diverse spectrum representations by employing multiple instances of scaled dot-product attention operating in parallel. The multi-head scaled dot-product attention comprises several attention heads. The final output from the multi-head attention is obtained by concatenating the outputs from all attention heads, where each attention head is computed independently, as outlined in Equations~\ref{equ4}, \ref{equ5}, and \ref{equ6}. Equation~\ref{equ5} linearly transforms the queries, keys, and vectors into multiple projections, referred to as attention heads. The attention is subsequently computed separately within each head, as depicted in Equation~\ref{equ4}. The outputs of these attention heads are then concatenated, as illustrated in Equation~\ref{equ6}.

\begin{equation}\label{equ4}
	  A^h(Q, K, V)  = \sum_{i=1}^{h} C_iW_i^o
\end{equation}

\begin{equation}\label{equ5}
	  C_i = A(QW_i^Q, KW_i^K, VW_i^V)
\end{equation}

\begin{equation}\label{equ6}
	 A(Q, K, V) = softmax(\frac{QK^T}{\sqrt{d_k}})V
\end{equation}

%\begin{figure}[ht]
%\centering
%\includegraphics[width=0.72\linewidth]{atten-layer.jpeg}
%\caption{Attention layer for concatenating the DIA spectra}
%\label{fig:attention-layer}
%\end{figure}  

\begin{table*}[htpb!]
\centering
\caption{Precision comparison of Transformer-DIA, DeepNovo-DIA and PepNet.}
\label{aba:tbl1}
\begin{tabular}{lcccccc}
\hline
\multicolumn{1}{l|}{\multirow{2}{*}{DIA Datasets}} & \multicolumn{3}{c|}{\textbf{Peptide-level performance}}     & \multicolumn{3}{c}{\textbf{Amino acid-level performance}} \\
\multicolumn{1}{l|}{}                              & Transformer-DIA   & DeepNovo-DIA & \multicolumn{1}{c|}{PepNet} & Transformer-DIA          & DeepNovo-DIA       & PepNet       \\ \hline
\multicolumn{1}{l|}{UTI}                           & \textbf{0.663} & 0.403        & \multicolumn{1}{c|}{0.296}  & \textbf{0.731}        & 0.566              & 0.396        \\
\multicolumn{1}{l|}{OC}                            & \textbf{0.798} & 0.417        & \multicolumn{1}{c|}{0.368}  & \textbf{0.829}        & 0.547              & 0.455        \\
\multicolumn{1}{l|}{Plasma}                        & \textbf{0.672} & 0.383        & \multicolumn{1}{c|}{0.351}  & \textbf{0.78}         & 0.702              & 0.509        \\ \hline
\multicolumn{7}{l}{{\footnotesize The best entry was in bold.}}                                                                                                                                                                                                                                                                                                                
\end{tabular}
\end{table*}

\subsection{Transformer Decoders in Transformer-DIA}

Transformer-DIA employs standard transformer encoders and decoders, as originally introduced by Vaswani et al.~\cite{vaswani2017attention} and used by Casanovo ~\cite{yilmaz2022novo}. Within this framework, the encoder's role is to capture latent representations of the input mass spectra, while the decoder is responsible for decoding the amino acid sequence associated with the originating peptide. Transformer-DIA's decoder takes three primary inputs: the spectrum representation vector generated by the transformer encoder, the previously predicted peptide sequence, and precursor information. Precursor information is represented as a tuple in the form of $(mass, charge)$ and is embedded into the model through sinusoidal functions~\cite{yilmaz2022novo}. This embedding process comprises two layers: one for embedding the mass vector and another for embedding the charge vector. These embeddings are subsequently summed to create the precursor embedding. Furthermore, the decoder accepts the preceding amino acids within the peptide sequence encoded by combining an amino acid embedding with a sinusoidal position embedding, representing their position within the sequence~\cite{yilmaz2022novo}.

%Transformer-DIA employs standard transformer encoders and decoders~\cite{vaswani2017attention}. The encoder is responsible for capturing latent representations of input mass spectra, while the decoder is tasked with decoding the amino acid sequence associated with the originating peptide. The Transformer-DIA decoder employs a transformer-based architecture for the reconstruction of amino acid sequences. It takes as input the spectrum representation vector generated by the transformer encoder, the previously predicted peptide sequence, and precursor information. The precursor information is represented as a tuple $(mass, charge)$ and is embedded using sinusoidal functions~\cite{yilmaz2022novo}. This embedding involves two layers: one for embedding the mass vector and another for embedding the charge vector. The resulting embeddings are then summed to create the precursor embedding. Additionally, the decoder encodes preceding amino acids in the peptide sequence by combining an amino acid embedding with a sinusoidal position embedding, which represents their position within the sequence~\cite{yilmaz2022novo}.

\subsection{Inference in Transformer-DIA}

To extract the probability distribution of amino acid sequences generated by the decoder output, we have employed a beam search algorithm. This algorithm operates iteratively at each decoding step, making predictions for the next amino acid based on the previous prediction, relevant precursor information, and the context derived from the spectrum's encoding through the transformer. Subsequently, it updates a score vector by identifying the top $k$ candidates with the highest scores, continuing this process until reaching either the stop token or reaching the maximum amino acid sequence length. The choice of $k$, determining the number of top candidates, can be adjusted as a hyperparameter. In our experiments, we set $k$ to 20. Transformer-DIA effectively filters out and terminates candidates exceeding the precursor mass tolerance. Ultimately, the algorithm selects the amino acid sequence with the highest confidence score as the prediction for each spectrum.  

\subsection{Model training in Transformer-DIA}\label{lab:GPU}

To train our model, we annotated each spectrum file with peptide sequences, sourced from the feature files by the peak caller from ref.~\cite{zhang2012peaks}. To ensure a fair comparison with DeepNovo-DIA and PepNet, all models underwent training on identical datasets. The datasets were randomly divided into training, validation, and testing sets with ratios of 0.90, 0.05, and 0.05, respectively. This division resulted in newly created datasets containing distinct peptides. We adopted cross-validation for the training and testing phases, leveraging a server equipped with 8 GeForce RTX 2080 GPUs. The vocabulary, consisting of 20 residues and incorporating a fixed modification for carbamidomethylation (C), was determined based on the datasets. Transformer-DIA underwent training using the same hyperparameters as Casanovo, which included an embedding size of 512, eight attention heads, a training batch size of 32, and a learning rate of $5 \times 10^{-4}$. Throughout the training process, we monitored the model's performance by calculating loss, peptide precision, amino acid precision, and recall after each epoch using the validation set. We optimized the model using the cross-entropy loss, minimizing the difference between the predicted peptide sequences and the ground truth peptide sequences.

%\begin{figure*}[t!]
 %\centering
 %   \begin{subfigure}{\textwidth}
  %  \centering
   %     \includegraphics[width=\linewidth]{oc-peptide-precision-rotation.png}
   %      \label{oc-peptide-precision}
  %  \end{subfigure}
 %   \begin{subfigure}{\textwidth}
   % \centering
    %    \includegraphics[width=\linewidth]{oc-aa-precision-rotation.png} 
   %      \label{oc-aa-precision}
   % \end{subfigure}
   % \begin{subfigure}{\textwidth}
   % \centering
     %   \includegraphics[width=\linewidth]{oc-aa-precision-rotation.png} 
     %    \label{oc-aa-precision}
   % \end{subfigure}
   %% \centering
     %   \includegraphics[width=\linewidth]{plasma-peptide-precision-rotation.png} 
    %    \label{plasma-peptide-precision}
    %\end{subfigure}
   % \begin{subfigure}{\textwidth}
   % \centering
    %    \includegraphics[width=\linewidth]{plasma-aa-precision-rotation.png} 
   %      \label{plasma-aa-precision}
   % \end{subfigure}
   % \begin{subfigure}{\textwidth}
   % \centering
   %     \includegraphics[width=\linewidth]{plasma-aa-recall-rotation.png} 
   %      \label{plasma-aa-recall}
   % \end{subfigure}
    
   % \caption{Performance of the Transformer-DIA variants on OC and Plasma data sets}
   % \label{plasma-oc-performance}
   
%\end{figure*}

\section{Data Sets}

We conducted a performance evaluation of Transformer-DIA in comparison to the state-of-the-art de novo sequencing methods, including Deepnovo-DIA and PepNet, using the DIA mass spectrometry datasets, as made available by Tran et al.~\cite{tran2019deep}. These datasets were obtained from the MassIVE repository, accessible under accession number MSV000082368 at~\url{https://massive.ucsd.edu/ProteoSAFe/dataset.jsp?task=88f95e8494cc4feeb3610e59f07f1d41}. Specifically, our training and evaluation utilized three distinct DIA datasets of Homo sapiens: urinary tract infection (UTI), ovarian cysts (OC), and plasma. These datasets contained 206,477 spectra, 203,780 spectra, and 1,097,400 spectra, respectively. Each dataset was randomly partitioned into separate training, validation, and testing sets with ratios of 0.9, 0.05, and 0.05, respectively. Furthermore, we employed a cross-validation approach, ensuring that no shared peptide sequences were present in the evaluation datasets. The DIA datasets were annotated by using the PEAKS DB \cite{zhang2012peaks}. The generated feature files contained the labels of ground truth peptide sequences.

%We evaluated the performance of Transformer-DIA compared with the state-of-the-art de novo sequencing model Deepnovo-DIA.  We used the same DIA mass spectrometry data sets released by \cite{tran2019deep} to train and test the. The dataset downloaded from the MassIVE repository under accession number MSV000082368 at \hyperref{https://massive.ucsd.edu/ProteoSAFe/dataset.jsp?task=88f95e8494cc4feeb3610e59f07f1d41}.  To train and evaluate the model, we used three homo sapien DIA data sets including urinary tract infection (UTI), ovarian cysts (OC), and plasma. The UTI, OC, and Plasma data sets consist of 206,477 spectra, 203,780 spectra, and 1097400 spectra respectively. Each data set is randomly partitioned into three training, validation, and testing sets with a ratio of 0.95, 0.5, and 0.5. We also applied cross-validation without having any shared peptide sequence.The released DIA data sets were annotated using the PEAKS DB database. The corresponding feature files contain grand truth peptide sequence as label acquired by PEAKS DB. The experimental results show that Transformer-DIA outperforms both DeepNovo-DIA and PepNet at both the amino acid level and significantly at the peptide level. 

\section{Experimental Results}

\subsection{Evaluation Metrics}

To evaluate the performance of Transformer-DIA, we employed metrics consistent with those utilized in DeepNovo-DIA~\cite{tran2019deep} and Casasnovo~\cite{yilmaz2022novo}. Amino acid precision assesses the ratio of correctly matched amino acids to the total predicted amino acids, while amino acid recall quantifies the ratio of correctly matched amino acids to the total number of amino acids present in the ground truth peptide sequences. A successful match is determined when the mass difference between predicted and ground truth amino acids falls below 0.1 Da, with any variations within a prefix or suffix not exceeding 0.5 Da. Peptide precision computes the ratio of fully correct predicted peptides to the total number of ground truth peptide sequences.

Additionally, we introduced the precision-coverage metric, where at the peptide level, coverage represents the ratio of predicted peptides to the original peptide count. In our experimental analyses, we conducted a comparative assessment of Transformer-DIA with DeepNovo-DIA and PepNet, given that both have consistently demonstrated superior performance compared to other de novo peptide sequencing tools designed for DIA data. Both DeepNovo-DIA and PepNet were trained and tested using parameters recommended by their respective authors.

\subsection{Performance comparison of Transformer-DIA variants}

\begin{figure*}[htpb] 
 \centering
    \begin{subfigure}{\textwidth}
    \centering
        \includegraphics[width=\linewidth]{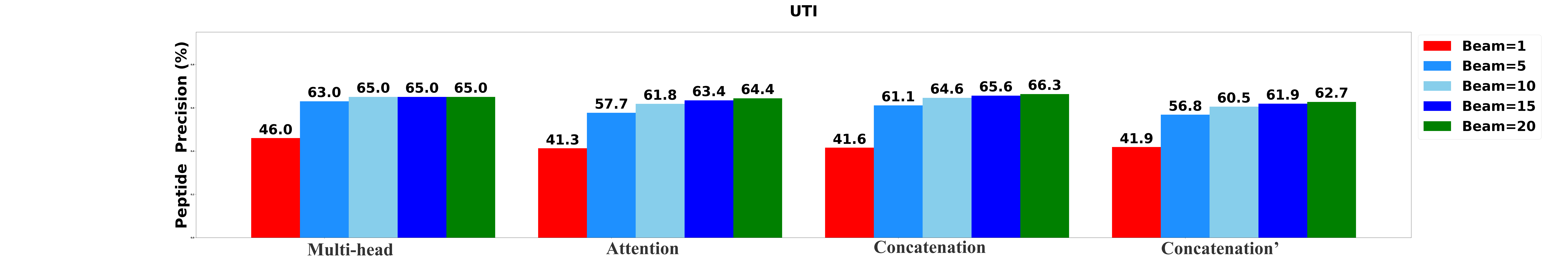} 
        \label{fig:timing31}
    \end{subfigure}
    \begin{subfigure}{\textwidth}
    \centering
        \includegraphics[width=\linewidth]{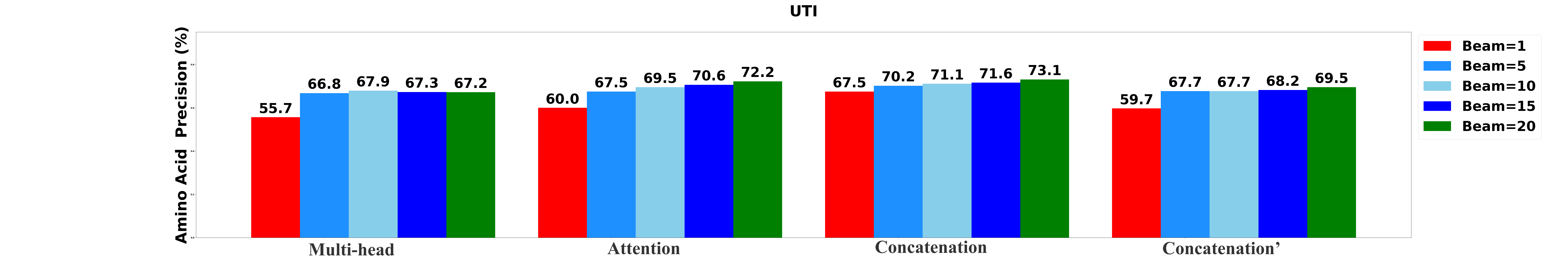} 
       \label{fig:timing32}
    \end{subfigure}
    \begin{subfigure}{\textwidth}
    \centering
        \includegraphics[width=\linewidth]{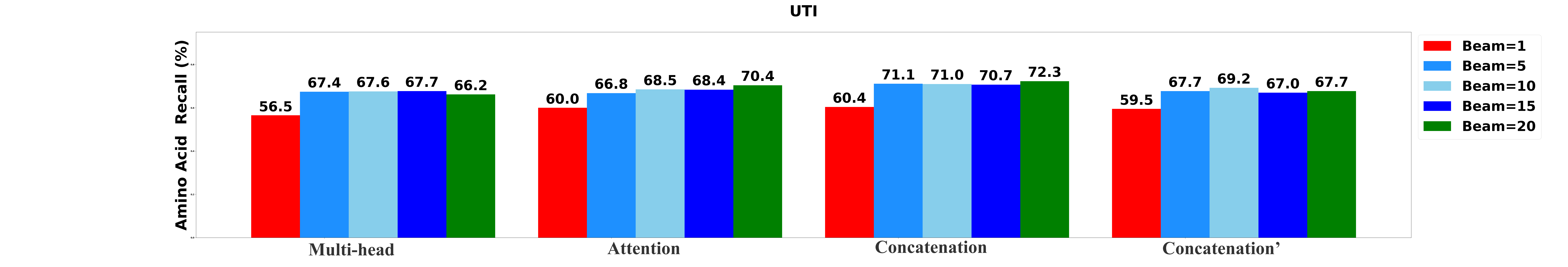} 
        \label{fig:timing33}
    \end{subfigure}
    \caption{Performance of the Transformer-DIA variants on UTI data set.}
    \label{UTI-performance}
\end{figure*}

\begin{figure*}[htpb] 
 \centering
    \begin{subfigure}{\textwidth}
    \centering
        \includegraphics[width=\linewidth]{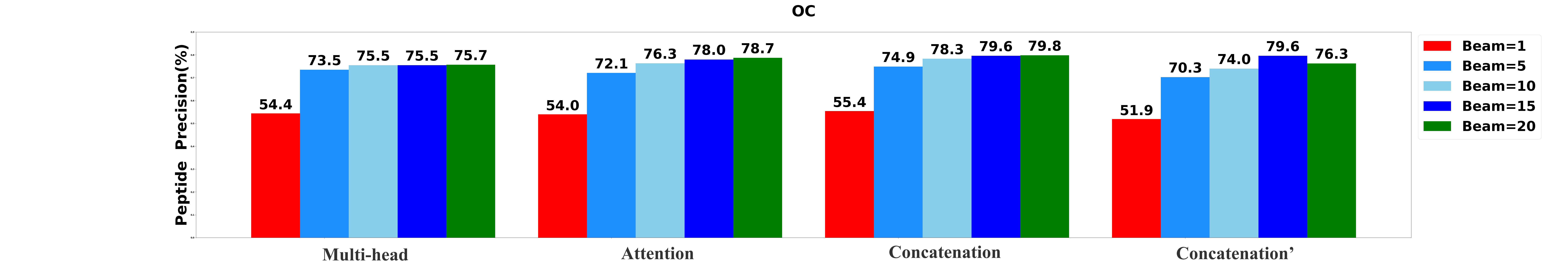} 
        \label{fig:timing34}
    \end{subfigure}
    \begin{subfigure}{\textwidth}
    \centering
        \includegraphics[width=\linewidth]{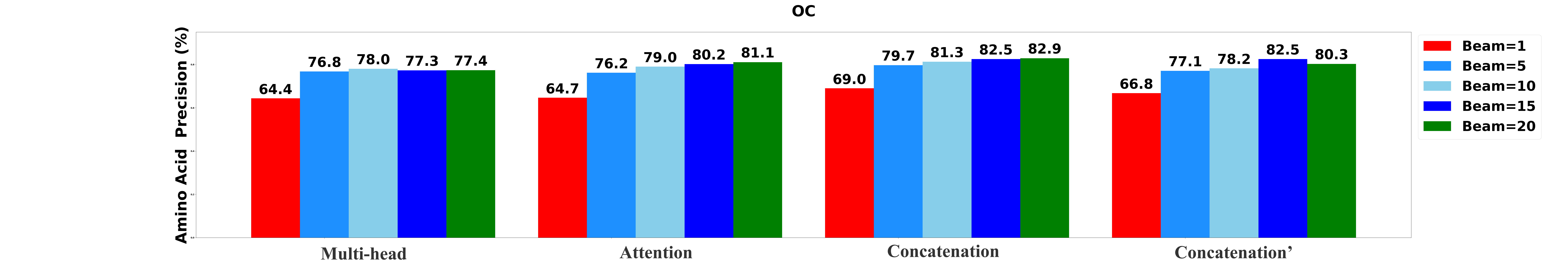} 
       \label{fig:timing35}
    \end{subfigure}
    \begin{subfigure}{\textwidth}
    \centering
        \includegraphics[width=\linewidth]{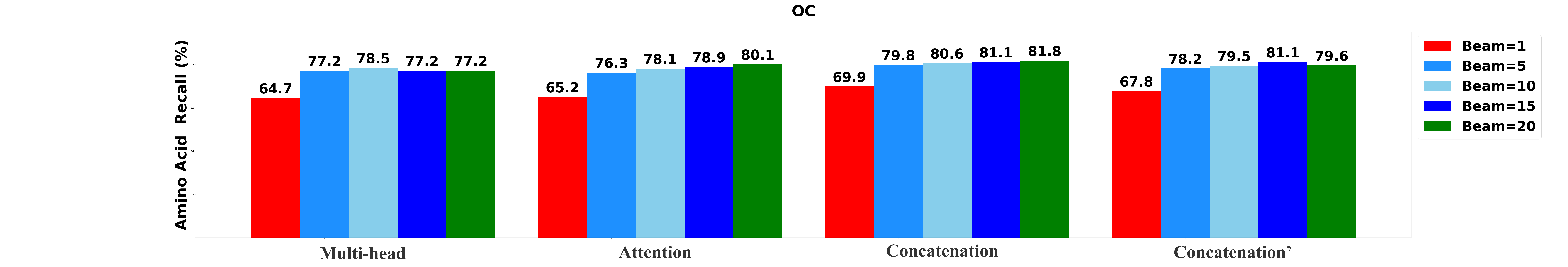} 
        \label{fig:timing36}
    \end{subfigure}
    \caption{Performance of the Transformer-DIA variants on OC data set.}
    \label{OC-performance}
\end{figure*}

\begin{figure*}[htpb] 
 \centering
    \begin{subfigure}{\textwidth}
    \centering
        \includegraphics[width=\linewidth]{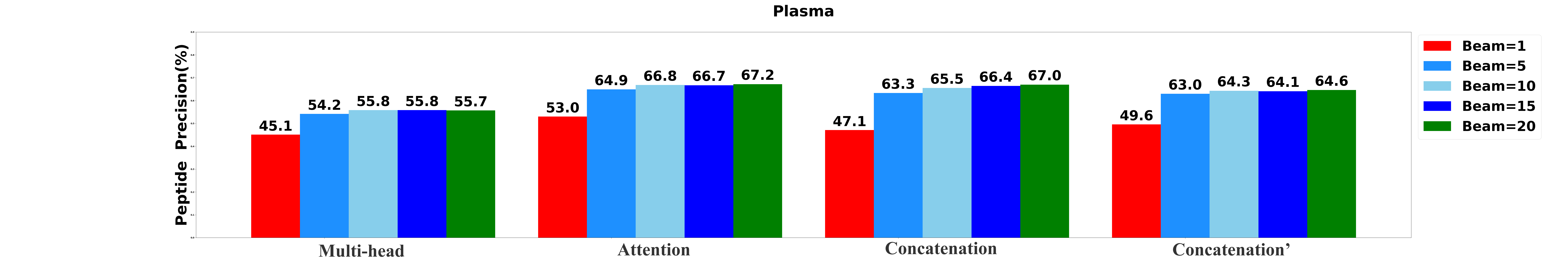} 
        \label{fig:timing37}
    \end{subfigure}
    \begin{subfigure}{\textwidth}
    \centering
        \includegraphics[width=\linewidth]{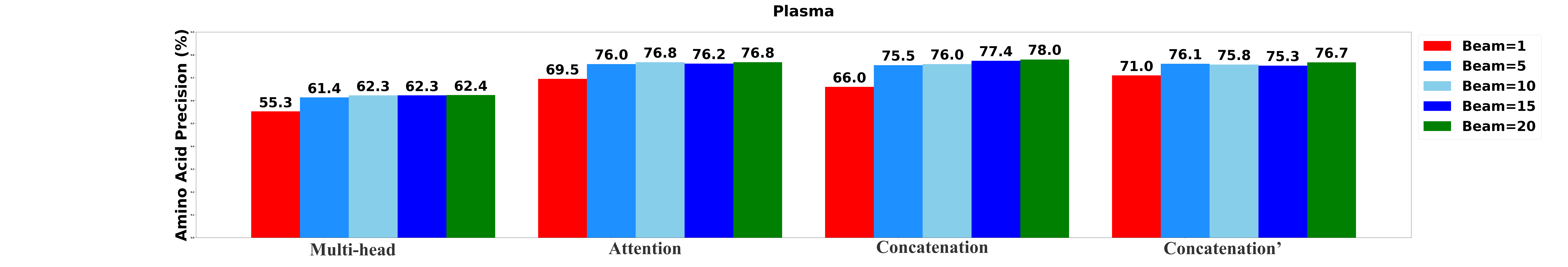} 
       \label{fig:timing38}
    \end{subfigure}
    \begin{subfigure}{\textwidth}
    \centering
        \includegraphics[width=\linewidth]{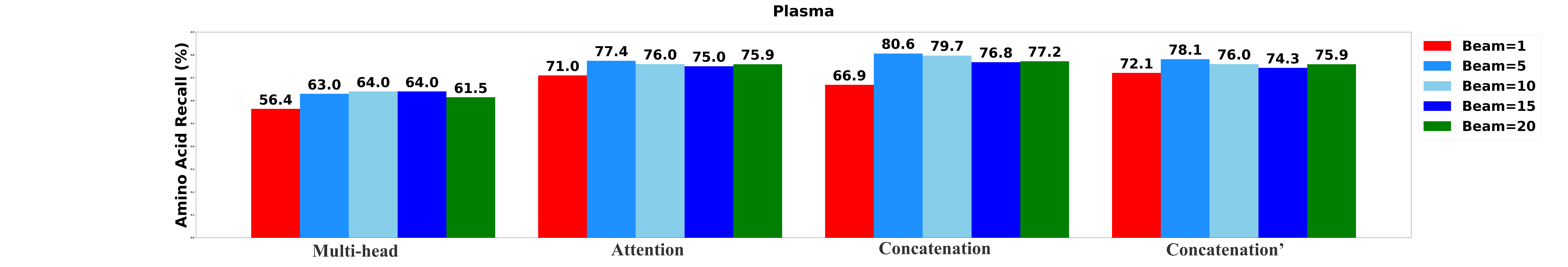} 
        \label{fig:timing39}
    \end{subfigure}
    \caption{Performance of the Transformer-DIA variants on Plasma data set.}
    \label{Plasma-performance}
\end{figure*}

To determine the most effective configurations, we developed four distinct versions of the Transformer-DIA model. The first variant, labeled as "Multi-head," incorporates a multi-head attention layer to integrate the encoded MS1 spectrum, MS2 spectra, and precursor profile. The second variant, labeled as "Attention," employs a standard attention layer to merge the encoded DIA spectra. The third variant, denoted as "Concatenation," integrates data using a concatenation layer, while the final variant, marked as "Concatenation$'$" utilizes a concatenation layer to integrate only MS2 spectra and precursor profile. We systematically assessed each variant of Transformer-DIA across five different beam sizes: 1, 5, 10, 15, and 20.

The results of the evaluation, including peptide precision, amino acid precision, and amino acid recall, are presented in Figs.\ref{UTI-performance},\ref{OC-performance}, and~\ref{Plasma-performance}. Generally, the highest performance across all model variants was consistently achieved when the beam size was set to 20. Among all the model versions, the "Concatenation" model consistently outperformed the others, closely followed by the "Attention" model. Conversely, the "Multi-head" model exhibited the least favorable performance when compared to the other variants, underscoring the significance of incorporating the MS1 spectrum.

%To identify the most effective configurations, we developed four distinct versions of the Transformer-DIA model. The first version, denoted as "Multi-head", utilizes a multi-head attention layer for the integration of encoded MS1 spectrum, MS2 spectra, and precursor profile. The second version, denoted as "Attention", employs a standard attention layer to fuse the encoded DIA spectra. The third version, denoted as "Concatenation", which incorporates a concatenation layer, and the last variant, denoted as "Concatenation'", uses a concatenation layer for integrating only MS2 spectra and precursor profile. We assessed every variant of Transformer-DIA using five different beam sizes: 1, 5, 10, 15, and 20, in addition to various combinations of MS1, MS2, and peaks data. The peptide precision, amino acid precision, and amino acid recall are shown in Figs.~\ref{UTI-performance},~\ref{OC-performance}, and~\ref{Plasma-performance}. In general, the highest performance was achieved when the beam size was set to 20 for all the model variants. Among all the model variants, the model "Concatenation" consistently achieving the highest overall performance followed by the model "Multi-head." The model "Concatenation’" yielded the least favorable performance when compared to the other versions, which highlight the importance of peaks data.

\subsection{Performance comparison of Transformer-DIA, DeepNovo-DIA and PepNet}

We evaluated the performance of Transformer-DIA using three distinct datasets: UTI, OC, and Plasma. The evaluation was conducted after 50,000 training iterations on a previously unseen test set. Our model was compared against two existing models specifically designed for DIA data: DeepNovo-DIA~\cite{tran2019deep} and PepNet~\cite{liu2022pepnet}. Table~\ref{aba:tbl1} presents the precision results at the peptide and amino acid levels for all three models across three benchmark cross-validation folds. Each fold's test set consists of spectra from a distinct species, with minimal overlap in the sets of peptides between species.

\begin{figure*}[htpb]
     \centering
     \begin{subfigure}[b]{0.32\textwidth}
         \centering
         \includegraphics[width=\textwidth]{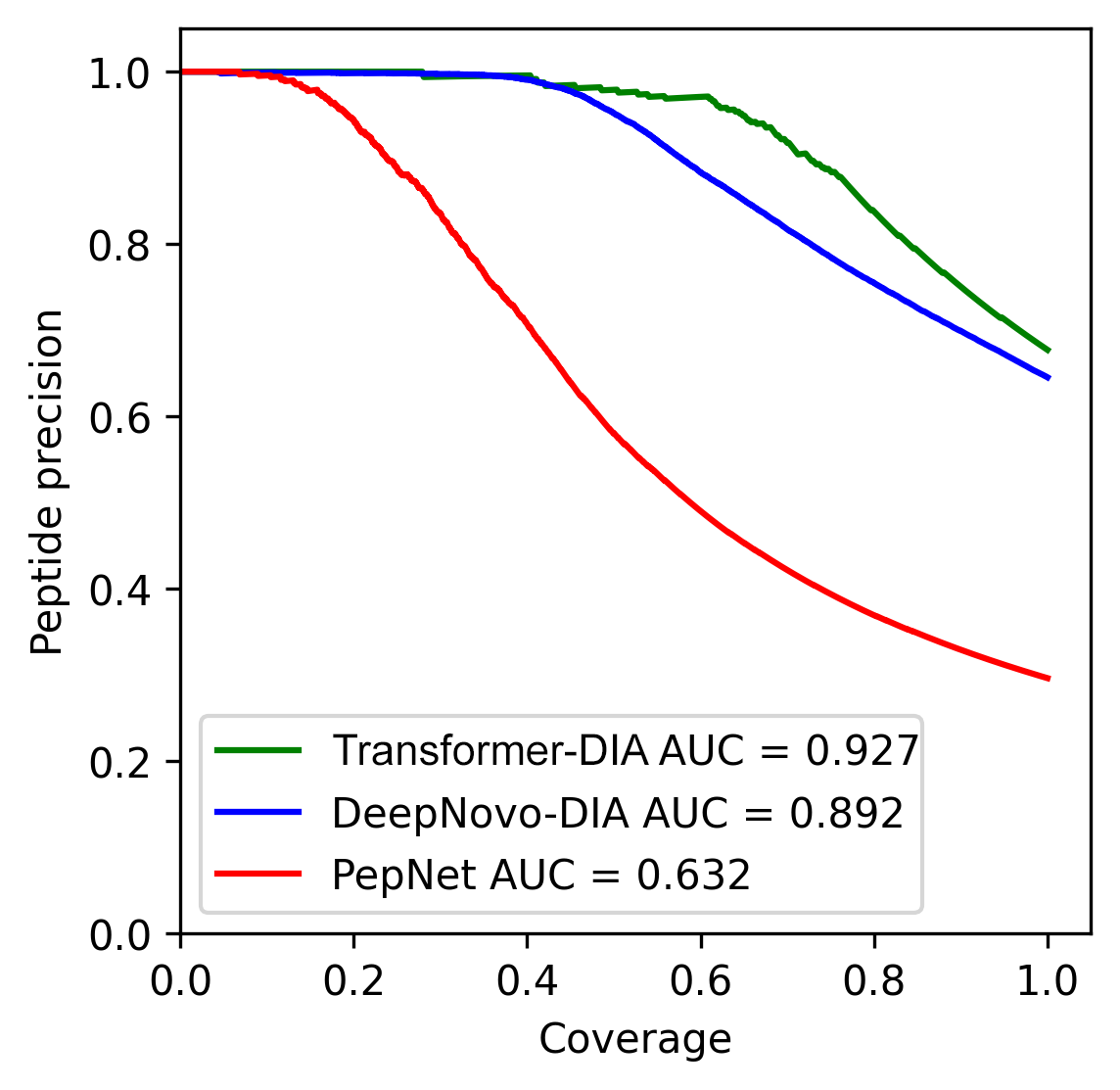}
          \caption{Precision-Coverage on UTI.}
         \label{fig:y equals x}
     \end{subfigure}
     \hfill
     \begin{subfigure}[b]{0.32\textwidth}
         \centering
         \includegraphics[width=\textwidth]{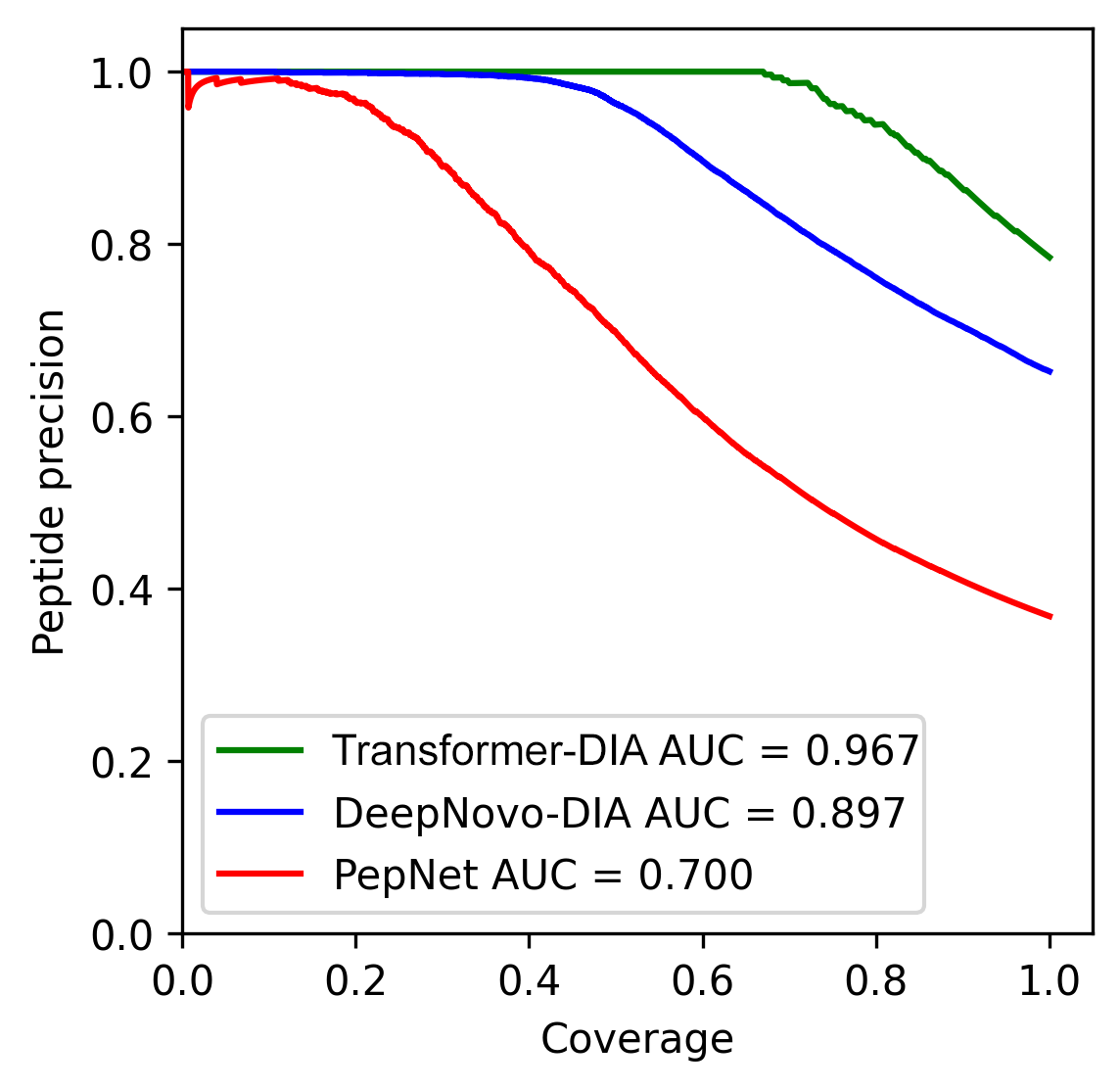}
        \caption{Precision-Coverage on OC.}
         \label{fig:three sin x}
     \end{subfigure}
     \begin{subfigure}[b]{0.32\textwidth}
         \centering
         \includegraphics[width=\textwidth]{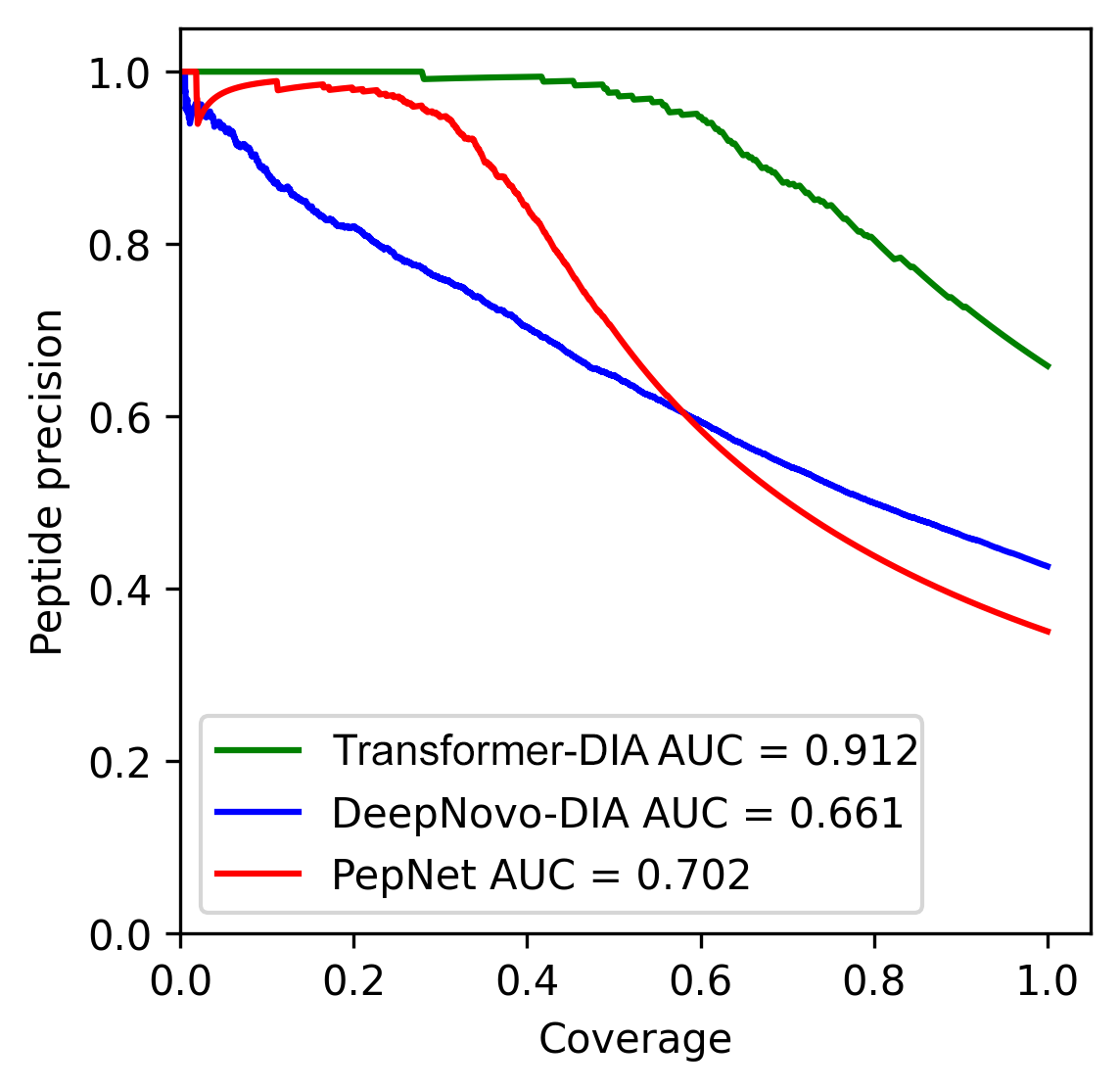}
         \caption{Precision-Coverage on Plasma.}
         \label{oc}
     \end{subfigure}
     \hfill
     
        \caption{Precision-coverage curves for Transformer-DIA, DeepNovo-DIA and PepNet.}
        \label{peptide-aa-precision}
\end{figure*}

The experimental results consistently demonstrate the superior performance of Transformer-DIA across all three datasets, both at the amino acid and peptide levels. Specifically, at the peptide level, Transformer-DIA improved peptide precision from 64.52\% to 75.46\% compared to DeepNovo-DIA, and from 91.45\% to 123.99\% compared to PepNet. At the amino acid level, Transformer-DIA exhibited a precision increase from 11.11\% to 63.07\% compared to DeepNovo-DIA, and from 53.24\% to 96.04\% compared to PepNet. These enhancements are visually represented in Fig.~\ref{peptide-aa-precision}, which displays the precision-coverage curves for peptides ranked by their confidence scores. These curves illustrate Transformer-DIA's superiority over DeepNovo-DIA and PepNet. Furthermore, when considering the area under the curve (AUC) metric, Transformer-DIA demonstrated an average improvement of 0.104 over DeepNovo-DIA and 0.256 over PepNet.

%We assessed the performance of Transformer-DIA using three distinct datasets: UTI, OC, and Plasma. Evaluation took place after 50,000 training iterations on a previously unseen test set. Our model was compared against two existing models tailored for DIA data: DeepNovo-DIA~\cite{tran2019deep} and PepNet~\cite{liu2022pepnet}. The Table~\ref{aba:tbl1} lists the peptide-level and amino acid-level precision of three competing models on all three benchmark cross-validation folds. Each fold’s test set contains spectra from a single  species, with a nearly disjoint sets of peptides between species. The experimental results consistently demonstrate Transformer-DIA's superiority across all three datasets, both at the amino acid and peptide levels. At the peptide level, Transformer-DIA improved peptide precision from 64.52\% to 75.46\% compared to DeepNovo-DIA and from 91.45\% to 123.99\% compared to PepNet. At the amino acid level, Transformer-DIA enhanced precision from 11.11\% to 63.07\% compared to DeepNovo-DIA and from 53.24\% to 96.04\% compared to PepNet. These improvements are visually illustrated in Fig.~\ref{peptide-aa-precision}, which presents the precision-coverage curves for peptides, ranked by their confidence scores. The curves clearly depict Transformer-DIA's superiority over DeepNovo-DIA and PepNet. Furthermore, when considering the area under the curve (AUC) metric, Transformer-DIA exhibited an average advantage of 0.104 over DeepNovo-DIA and 0.256 over PepNet.

%We evaluated the performance of Transformer-DIA on the three data sets, i.e., UTI, OC, and Plasma. During the evaluation step, we evaluated the performance of the trained model after 50,000 iterations on the test set including unseen peptide sequences. We also compared our model to two models developed for DIA data sest including DeepNovo-DIA \cite{tran2019deep} and PepNet \cite{liu2022pepnet}. The experimental results show that Transformer-DIA outperforms DeepNovo-DIA and PepNet for all three datasets at both amino acid and peptide levels. Both DeepNovo-DIA and PepNet trained and tested on the same data sets. As we can see at the table\ref{aba:tbl1}, Transformer-DIA achieved significant improvement at the peptide level in comparison with two other models. It improved the peptide precision 64.52\% to 75.46\% and 91.45\% to 123.99\% compare to the DeepNovo-DIA and PepNet respectively. At the amino acid level, Transformer-DIA improved the precision 11.11\% to 63.07\% and 53.24\% to 96.04\% compare to the DeepNovo-DIA and PepNet respectively. Figure \ref{peptide-aa-precision} illustrates precision-coverage curve at peptide level sorted predicted peptides based on their confidence scores. The Figure \ref{peptide-aa-precision} curves also exhibit that Transformer-DIA outperforms DeepNovo-DIA and PepNet. It becomes evident that the area under the curve (AUC) metric displays an average superiority of 0.104 over DeepNovo-DIA and  0.256 over PepNet.

\section{Discussion and Conclusion}

In order to assess the runtime performance of our extended model in comparison to the original Casanova, we conducted an experiment where we annotated the spectra files of each dataset and evaluated them on the same test set. We used the same server as described in Section~\ref{lab:GPU}. The results of this analysis indicate that the training process of the extended model is nearly as efficient as the original Casanova. More specifically, training the Casanova model took 6 hours for the plasma dataset, 9 hours for the OC dataset, and 11 hours for the UTI dataset. On the contrary, to complete the training of the Casanova-DIA model, 20 hours for the plasma dataset, 15 hours for the OC dataset, and 17 hours for the UTI dataset are required. This finding suggests that the extended model's runtime performance remains within acceptable bounds.

The development of various Casanova-DIA versions, featuring increased beam sizes and diverse architectures, highlights the model's capability to achieve superior performance at the peptide level while ensuring scalability. This scalability translates to minimal impact on runtime, making Casanova-DIA a practical choice for real-world applications.

In conclusion, our study demonstrates that Casanova-DIA, as an extension of the Casanova model, significantly improves the performance of de novo peptide sequencing for DIA data, particularly at the challenging peptide level. This enhancement in performance, coupled with its scalability and comparable runtime efficiency, positions Casanova-DIA as a promising tool for advancing peptide sequencing in the field of LC-MS/MS DIA data analysis. In our upcoming work, we intend to enhance the model's capabilities for interpreting DIA spectra.

%To assess runtime performance in comparison to Casanova, we annotated the spectra files of each dataset and evaluated them on the same test set. The results reveal that the extended model's training process is almost as efficient as the original Casanova (Data not shown here), which is deemed acceptable.

%The development of various Casanova-DIA versions, including those with increased beam size and different architectures, underscores the model's ability to achieve superior performance at the peptide level while maintaining scalability, ensuring minimal impact on runtime. Compared to the state-of-the-art model designed for DIA spectra, Casanova-DIA, as an extension of Casanova, demonstrates a significant enhancement in performance, particularly at the challenging peptide level. 

%\section*{Acknowledgment}
%This work was supported by the National Library of Medicine of the National Institutes of Health under award number R15LM013460, and the National Center for Complementary \&
%Integrative Health of the National Institutes of Health under the award number R01AT011618 https://www.nih.gov/.

%\section*{Acknowledgment}
\bibliographystyle{./IEEEtran}
\bibliography{./conference_101719}
\FloatBarrier
\end{document}